\documentclass[journal]{IEEEtran}
\usepackage{amsmath,amssymb,amsfonts}
\usepackage{graphicx}
\usepackage{algorithm}
\usepackage{algorithmic}
\usepackage{array}
\usepackage{textcomp}
\usepackage{caption}
\usepackage{subcaption}
\usepackage{cite}
\usepackage{bm}
\usepackage{balance}
\usepackage[usenames, dvipsnames]{color}

\newcommand{\vect}[1]{\boldsymbol{#1}}
\newcommand{\mat}[1]{\mathbf{#1}}

\begin{document}
	
	\title{Conical Localization via Modified Polar Representation: A Unified Framework for Robust 3-D Positioning with 1-D Sensor Arrays}
	
	\author{Ehsan Alamdari, Rouhollah Amiri
	\thanks{The authors are with Department of Electrical Engineering, Sharif University of Technology, Tehran, Iran, (e-mail: amiri@sharif.edu.)}}
	
	\maketitle
	
	\begin{abstract}
		This paper presents a unified framework for robust three-dimensional (3-D) source localization using a network of sensors equipped with one-dimensional (1-D) linear arrays. While such arrays offer practical advantages in terms of cost and size, existing localization methods suffer from a fundamental limitation: their performance degrades significantly as the source moves into the far-field, a common challenge known as the thresholding effect. To address this issue, we reformulate the localization problem in the modified polar representation (MPR) coordinate system, which parameterizes the source location using its azimuth, elevation, and inverse-range. We have developed a constrained weighted least squares (CWLS) estimator, which is subsequently transformed into a tight semidefinite programming (SDP) problem via semidefinite relaxation, enhanced with additional constraints to improve accuracy. Simulation results demonstrate that the proposed estimator attains the Cramér-Rao lower bound (CRLB) for both angle and inverse-range estimation in near-field scenarios. More importantly, it maintains this optimal performance in the far-field, substantially outperforming state-of-the-art methods, which exhibit significant error at large ranges. The proposed solution thus provides a reliable, unified localization system that is effective irrespective of the source range.
	\end{abstract}
	
	
	\section{Introduction}
	\IEEEPARstart{P}{assive} source localization is a widely studied topic among researchers due to its numerous applications in various fields, including sonar, radar, 5G networks, and electronic warfare \cite{poisel2012, witrisal2016, amiri2018, sadeghi2021, shamsian2025, sun2024, wang2015, wang2018, amiri2017, zhou2024}. Two primary approaches dominate this research area. The time-based measurements approach relies on temporal parameters, such as time of arrival (TOA), Time difference of Arrival (TDOA) or signal phase differences. These methods typically require precise time synchronization among sensors and knowledge of the signal propagation speed (e.g., the speed of light or sound). By leveraging these measurements, the source's position can be estimated. Although this approach is highly effective in applications like global navigation satellite system (GNSS) or localization in wireless sensor networks, the demand for accurate synchronization can pose challenges in complex environments.
	
	The angle-based measurements approach focuses on determining the Angle of Arrival (AOA) of the signal, typically using antenna arrays or directional sensors \cite{wang2015, wang2018, amiri2017, zhou2024}. While traditional three-dimensional localization, reliant on well-established azimuth-elevation angle measurements which necessitate a planar array for implementation, is widely recognized, recent years have witnessed a rapidly growing interest in source localization utilizing linear arrays. One-dimensional antenna arrays require fewer antenna elements and, consequently, less backend hardware compared to two-dimensional arrays, resulting in lower manufacturing costs and more compact designs. This advantage is particularly significant in applications where spatial constraints limit the placement of antenna arrays. In contrast to the two-dimensional arrays, which provides outputs in terms of azimuth and elevation angles, thereby defining a line connecting each sensor and the source, in a one-dimensional array, the locus of points where the source may be located, forms a conical surface. The axis of this cone aligns with the arrays orientation vector and its vertex corresponds to the sensor's position. By utilizing a network of such sensors equipped with linear arrays, each defining a conical surface, the position of the source may be determined by intersecting these cones.
	
	\subsection{Related Works}
	Several publications have appeared in recent years addressing the problem of 3-D source localization using a network of linear arrays and various approaches have been employed to solve this problem \cite{zou2020a, zou2020b, alamdari2022, sun2022, sun2023, chen2023, zou2025, li2025}.
	The early studies \cite{zou2020a} and \cite{zou2020b} by Zou et al. represents the initial efforts undertaken in this field. In \cite{zou2020a} the localization problem is solved using an iterative method and \cite{zou2020b} formulates it as a constrained weighted least squares (CWLS) problem and solves the non-convex problem by applying semidefinite relaxation (SDR) technique. A key limitation of \cite{zou2020a, zou2020b} is that they do not address the problem in the most general case and rely on specific assumption regarding the placement of sensors and their attitude vectors.
	
	To address this issue, \cite{alamdari2022, sun2022} tried to solve the problem without imposing specific geometric condition. \cite{alamdari2022} and \cite{sun2022} both formulate the location finding problem as a CWLS problem but each employs a distinct SDR techniques to recast it as a convex optimization problem. In \cite{sun2023, chen2023, zou2025}, the solution to this problem was obtained employing iterative techniques. To best of our knowledge, \cite{li2025} is the only work that has addressed this problem algebraically with satisfactory performance. Nevertheless, its limitation lies in the fact that the problem has not been solved in the case of minimum sensors configuration.
	
	Point source localization is only meaningful when the distance between the source and the sensors is not excessively large. Under conditions where this distance is significantly extended, even minor angular measurement error in the sensors will cause the intersection point of the conical surfaces to deviate considerably from the true source location. Under this condition, the estimation error is so significant that it invalidates the results, rendering them essentially meaningless and unsuitable for practical application.
	
	This issue also manifests in TDOA-based localization, where hyperbolic surfaces become nearly parallel at large distances from the sensors \cite{wang2017, tang2025, sun2019}. Consequently, even minor measurement errors can cause a significant displacement of their intersection point, a phenomenon known in the literature as the threshold effect. An innovative solution to overcome this challenge was first introduced in \cite{wang2017}, which proposed a new coordinate system called the modified polar representation (MPR) to provide reliable localization in a unified manner. This system is similar to spherical coordinates but substitutes the range with its reciprocal which is called the inverse-range. When developing an estimator based on this coordinate system, a unified framework for localization is established, applicable to both near and far fields. Specifically, when the source is at a close distance, point localization is achieved by estimating the angle and inverse range. Conversely, when the source is at a significantly distant range, although the inverse range estimate converges to zero, the angle estimates remain valid, and the corresponding Direction of Arrival (DOA) estimation for the source remains reliable.
	
	\subsection{Motivation and Contributions}
	Although source localization using linear arrays is considered a relatively recent topic, multiple estimators have been proposed that can accurately determine the source location under mild measurement noise. However, the previous studies do not take into account the scenarios where the source location is extensively far from the sensors.
	In this paper, the source location finding problem using 1-D arrays is reformulated in the MPR coordinate system to develop a unified near- and far-field estimator using angle measurements. The proposed method addresses the location finding problem by first formulating it as a non-convex CWLS problem and then recasting it into an SDP problem via Semidefinite Relaxation (SDR). Simulation results indicates that the proposed method can reach the Cramer-Rao lower bound (CRLB) performance in both near-field and far-field scenarios and has no limitation in the sensors configuration.
	
	\subsection{Notations}
	The following notations will be used throughout the paper. Bold lower- and upper-case letters denote column vectors and matrices. The subscripts $(\cdot)^{T}$ and $(\cdot)^{-1}$ stand for transpose and inverse operations, respectively. $\vect{a}^{\circ}$ is the true value of $\vect{a}$ when it is contaminated with noise. $\text{diag}(\vect{a})$ is a diagonal matrix with elements of $\vect{a}$ on the diagonal. $\text{tr}\{\mat{A}\}$ is the trace of matrix $\mat{A}$. $[\vect{a}]_i$, $[\mat{A}]_{i,j}$, and $[\mat{A}]_{i,:}$ are the $i$th element of $\vect{a}$, the $(i,j)$th element of $\mat{A}$ and the $i$th row of $\mat{A}$ respectively. $[\mat{A}]_{i:j,k}$ is a vector formed by the elements at row $i$ to $j$ in column $k$ of matrix $\mat{A}$. $\vect{0}_{k,l}$, and $\mat{I}_{N}$ represent the $k \times l$ all-zero matrix and identity matrix of size $N$, respectively. $\mat{A} \succcurlyeq 0$ indicates that matrix $\mat{A}$ is positive semi-definite and $\| \cdot \|$ is the Euclidean norm.
	
	\section{Problem Statement}
	Let us begin with the general case of 3-D localization of a source located at unknown location $\vect{u}^{o}$ using an array of $N$ sensors whose location and attitude vectors is known and denoted respectively by $\vect{s}_{i}^{o}$ and $\vect{a}_{i}^{o}$, $i=1,2,\ldots,N$. Each sensor is equipped with a small directional array which is capable of measuring the 1-D angle from the source to the sensor.
	
	The observed 1-D angle in the $i$th sensor can be defined as
	\begin{equation}\label{eq:meas}
	\psi_{i} = \psi_{i}^{o} + n_{i}
	\end{equation}
	where
	\begin{equation}\label{eq:Angle}
	\psi_{i}^{o} = \cos^{-1}\left( \frac{{\vect{a}_{i}^{o}}^{T}\left( \vect{u}^{o} - \vect{s}_{i}^{o} \right)}{\left\| \vect{u}^{o} - \vect{s}_{i}^{o} \right\|} \right)
	\end{equation}
	is the true angle and $n_{i}$ is the measurement noise. Collecting all the $N$ measurements in vector form gives
	\begin{equation}
	\vect{\psi} = \vect{\psi}^{o} + \vect{n}
	\end{equation}
	where $\vect{\psi}^{o} = [\psi_{1}^{o},\psi_{2}^{o},\ldots,\psi_{N}^{o}]^{T}$ and $\vect{\psi}$, $\vect{n}$ are defined similarly. It is reasonable to model the noise vector $\vect{n}$ as a zero mean Gaussian random vector with covariance matrix $\mat{Q}$.
	
	The above equations represent the problem in cartesian coordinate. In order to develop a unified estimator for both near field and far field sources, we shall express the unknown source parameters in MPR coordinate as follows \cite{wang2017}.
	
	Let $\tilde{\vect{u}}^{o} = [\varphi^{o},\theta^{o},g^{o}]^{T}$ where $\varphi^{o}$ and $\theta^{o}$ are the true azimuth and elevation angles of the source, respectively. $g^{o}$ is the inverse-range of the source and can be defined as
	\begin{equation}
	g^{o} = \frac{1}{r^{o}}
	\end{equation}
	In addition, let us define ${\vect{\rho}}^{o}$ as a unit vector, pointing form the origin to the source location as
	\begin{equation}
	{\vect{\rho}}^{o} = \frac{\vect{u}^{o}}{r^{o}}
	\end{equation}
	Taking cosine on both sides of \eqref{eq:Angle} and multiplying the numerator and denominator of the right side by $g^{o}$ gives:
	\begin{equation} \label{eq:Angle_Cos}
	\cos\left( \psi_{i}^{o} \right) = \frac{{\vect{a}_{i}^{o}}^{T}\left( {\vect{\rho}}^{o} - \vect{s}_{i}^{o}g^{o} \right)}{\left\| {\vect{\rho}}^{o} - \vect{s}_{i}^{o}g^{o} \right\|}
	\end{equation}
	By leveraging \eqref{eq:meas} and under small noise condition where $\sin(n_{i}) = n_{i}$ and $\cos(n_{i}) = 1$ we have
	\begin{equation}\label{eq:noisy_cos}
	\cos\left( \psi_{i}^{o} \right) = \cos\left( \psi_{i} \right) + \sin\left( \psi_{i} \right)n_{i}
	\end{equation}
	Inserting \eqref{eq:noisy_cos} into \eqref{eq:Angle_Cos} and some simple algebraic manipulation yields in
	\begin{align}\label{eq:psude-linear}
	&{\vect{a}_{i}^{o}}^{T}\vect{s}_{i}^{o}g^{o} - {\vect{a}_{i}^{o}}^{T}{\vect{\rho}}^{o} + \cos\left( \psi_{i} \right)\left\| {\vect{\rho}}^{o} - \vect{s}_{i}^{o}g^{o} \right\|\nonumber\\
	&= - \sin\left( \psi_{i} \right)\left\| {\vect{\rho}}^{o} - \vect{s}_{i}^{o}g^{o} \right\| n_{i}, i=1,2,\ldots,N
	\end{align}
	
	By defining $\vect{h} = [g^{o}, {\vect{\rho}}^{oT}, r_{1}^{o}, \ldots, r_{N}^{o}]^{T}$ with $r_{i}^{o} = \left\| {\vect{\rho}}^{o} - \vect{s}_{i}^{o}g^{o} \right\|$ for $i = 1,\ldots,N$, \eqref{eq:psude-linear} can be written in matrix form as
	\begin{equation}\label{eq:matrix_equation}
	\mat{F}\vect{h} = \mat{B}\vect{n}
	\end{equation}
	where
	\begin{subequations}
	\begin{align}
	\mat{F} &= [\mat{F}_{1}, \mat{F}_{2}] \\
	[\mat{F}_{1}]_{i,:} &= [{\vect{a}_{i}^{o}}^{T}\vect{s}_{i}^{o}, -{\vect{a}_{i}^{o}}^{T}], i=1,\ldots,N \\
	\mat{F}_{2} &= \text{diag}\left( [\cos(\psi_{1}),\cos(\psi_{2}),\ldots,\cos(\psi_{N})]^{T} \right)
	\end{align}
	\end{subequations}
	and $\mat{B}$ is given by
	\begin{equation}
	\mat{B} = -\text{diag}\left( [r_1^o\sin(\psi_{1}),\ldots,r_N^o\sin(\psi_{N}) ]^{T} \right)
	\end{equation}
	
	Now we can derive a constrained WLS optimization problem according to \eqref{eq:matrix_equation} as
	\begin{subequations}\label{eq:Problem1}
	\begin{align}
	\min_{\vect{h}} &\left( \mat{F}\vect{h} \right)^{T}\mat{W}\left( \mat{F}\vect{h} \right)\nonumber\\
	\text{s.t.}\,\,
		&r_{i}^{o} = \left\| {\vect{\rho}}^{o} - \vect{s}_{i}^{o}g^{o} \right\|, i=1,2,\ldots,N \\
	&\left\| {\vect{\rho}}^{o} \right\| = 1
	\end{align}	
\end{subequations}
	where $\mat{W} = \left( \mat{B}\mat{Q}\mat{B}^{T} \right)^{-1}$ represents the weighting matrix.
	
	Let us define the matrix $\mat{H} = \vect{h}\vect{h}^{T}$, the CSWL problem \eqref{eq:Problem1} can be rewritten as
	\begin{subequations}\label{eq:Problem2}
		\begin{align}
		&\min_{\mat{H}} \text{tr}\left\{ \mat{F}^{T}\mat{W}\mat{F}\mat{H} \right\}\nonumber\\
			&\text{s.t.}\,\,
		\text{tr}\left\{ [\mat{H}]_{2:4,2:4} \right\} = 1 \\
		&\text{tr}\left\{ \mat{C}_{i}\mat{C}_{i}^{T}\mat{H} \right\} = [\mat{H}]_{i+4,i+4}, i=1,2,\ldots,N \\
		&\mat{H} \succcurlyeq 0 \\
		&\text{rank}\left\{ \mat{H} \right\} = 1	
		\end{align}
	\end{subequations}
	where $\mat{C}_{i} = [-\vect{s}_{i}^{o},\mat{I}_{3},\vect{0}_{3,N}]$.
	
	Although problem \eqref{eq:Problem2} has an affine objective function in terms of the elements of $\mat{H}$, the rank one constraint makes it a non-convex optimization problem. To overcome this issue, we simply drop the rank constraint to obtain a convex SDP.
	\begin{subequations}\label{eq:Problem3}
	\begin{align}
	&\min_{\mat{H}} \text{tr}\left\{ \mat{F}^{T}\mat{W}\mat{F}\mat{H} \right\}\nonumber\\
	&\text{s.t.}\,\,
		\text{tr}\left\{ [\mat{H}]_{2:4,2:4} \right\} = 1 \label{eq:Problem3_Const1}\\
	&\text{tr}\left\{ \mat{C}_{i}\mat{C}_{i}^{T}\mat{H} \right\} = [\mat{H}]_{i+4,i+4}, i=1,2,\ldots,N \label{eq:Problem3_Const2}\\
	&\mat{H} \succcurlyeq 0 \label{eq:Problem3_Const3}
	\end{align}		
	\end{subequations}
	
	The above problem is convex but simulation shows that, in most cases, it converges to an inaccurate solution with a rank higher that one, indicating that the solution to \eqref{eq:Problem3} is not the optimal solution to the original problem \eqref{eq:Problem1}. To reach a more accurate solution, some reasonable constraint could be employed to tighten the feasible region of the SDP problem.
	
	Let ${\mat{H}}^*$ be located in the feasible region of \eqref{eq:Problem3}, i.e., satisfying \eqref{eq:Problem3_Const1}-\eqref{eq:Problem3_Const3}. Generally, ${\mat{H}}^*$ is not of rank one, hence there exists vector ${\vect{h}}^* = [{g}^*,{\vect{\rho}}^{*T},{r_{1}}^*,\ldots,{r_{N}}^*]^{T}$ where ${\mat{H}}^* \succcurlyeq {\vect{h}}^*{\vect{h}}^{*T}$. From \eqref{eq:Problem3_Const2} we have
	\begin{align}
	[{\mat{H}}^*]_{i+4,i+4} = \text{tr}\left\{ \mat{C}_{i}\mat{C}_{i}^{T}{\mat{H}}^* \right\} &\geq \text{tr}\left\{ \mat{C}_{i}\mat{C}_{i}^{T}{\vect{h}^*}{\vect{h}}^{*T} \right\} \nonumber\\&= \left\| {\vect{\rho}}^* - \vect{s}_{i}^{o}{g}^* \right\|^{2}
	\end{align}
	Accordingly, we may deduce that in the feasible region of the problem \eqref{eq:Problem3} we always have
	\begin{equation}\label{eq:Feasible_Region1}
	\left\| \vect{\rho}^* - \vect{s}_{i}^{o}g^* \right\| \leq r_{i}^*
	\end{equation}
	Multiplying both sides of \eqref{eq:Feasible_Region1} by the positive value $g^*$ gives
	\begin{equation}
	\left\| \vect{\rho}^*g^* - \vect{s}_{i}^{o}g^{*2} \right\| \leq r_{i}^*g^*
	\end{equation}
	which can be written as
	\begin{equation}\label{eq:tighten_const1}
	\left\| [\mat{H}]_{2:4,1} - \vect{s}_{i}^{o}[\mat{H}]_{1,1} \right\| \leq [\mat{H}]_{4+i,1}, i=1,2,\ldots,N
	\end{equation}
	A similar approach can be applied by multiplying the positive value $r_{j}^*$ to the both sides of \eqref{eq:Feasible_Region1} as
	\begin{equation}
	\left\| \vect{\rho}^*r_{j}^* - \vect{s}_{i}^{o}g^*r_{j}^* \right\| \leq r_{i}^*r_{j}^*
	\end{equation}
	or equivalently
	\begin{align}\label{eq:tighten_const2}
	&\left\| [\mat{H}]_{2:4,4+j} - \vect{s}_{i}^{o}[\mat{H}]_{1,4+j} \right\| \leq [\mat{H}]_{4+j,4+i},\nonumber\\
	&i=1,2,\ldots,N, j=1,2,\ldots,N
	\end{align}
	Analogously, multiplying $[\vect{u}]_{j}$ to the both sides of \eqref{eq:Feasible_Region1} gives
	\begin{align}\label{eq:tighten_const3}
	&\left\| [\mat{H}]_{2:4,1+j} - \vect{s}_{i}^{o}[\mat{H}]_{1,1+j} \right\| \leq \text{sign}\left( [\vect{u}]_{j} \right)[\mat{H}]_{1+j,4+i},\nonumber\\
	&i=1,2,\ldots,N, j=1,2,3
	\end{align}
	
	By adding extra constraints in \eqref{eq:tighten_const1}, \eqref{eq:tighten_const2}, and \eqref{eq:tighten_const3}, we can tighten the feasible region of the SDP problem \eqref{eq:Problem3} and reach a more accurate solution. As a result, we have
	\begin{align}\label{eq:Problem4}
	&\min_{\mat{H}} \text{tr}\left\{ \mat{F}^{T}\mat{W}\mat{F}\mat{H} \right\}\nonumber\\
	&\text{s.t. } \eqref{eq:Problem3_Const1},\eqref{eq:Problem3_Const2},\eqref{eq:Problem3_Const3},\eqref{eq:tighten_const1}, \eqref{eq:tighten_const2}, \eqref{eq:tighten_const3}
	\end{align}
	
	Let $\mat{H}^{*}$ be the optimal solution to the SDP problem \eqref{eq:Problem4}. By applying eigen-value decomposition to $\mat{H}^{*}$ we can derive $\vect{h}^{*}$ as the eigenvector which has the largest eigenvalue $\lambda^{*}$. Consequently, we can recover the estimated source position in MPR as
	\begin{subequations}
	\begin{align}
	\widehat{g} &= [\vect{h}^{*}]_{1} \\
	\widehat{\varphi} &= \text{arctan2}\left( [\vect{h}^{*}]_{3},[\vect{h}^{*}]_{2} \right) \\
	\widehat{\theta} &= \text{arctan2}\left( [\vect{h}^{*}]_{4},\sqrt{\left( [\vect{h}^{*}]_{2} \right)^{2} + \left( [\vect{h}^{*}]_{3} \right)^{2}} \right)
	\end{align}
	\end{subequations}
	
	\textit{Remark 1:} Since the sign of $[\vect{u}]_{j}$ is unknown beforehand, a practical approach involves initially solving problem \eqref{eq:Problem4} without enforcing constraints \eqref{eq:tighten_const3} to ascertain the correct signs. Based on the results, the constraints \eqref{eq:tighten_const3} can then be properly formulated and incorporated into the final solution of problem \eqref{eq:Problem4}.
	
	\textit{Remark 2:} The weighting matrix $\mat{W}$ is a function of the unknown source position, as it depends on the matrix $\mat{B}$. To address this circular dependency, a two-step iterative procedure can be employed. First, an initial solution is obtained by solving the problem with an approximate weighting matrix, typically set to $\mat{W} = \mat{Q}^{-1}$. The result of this step provides a preliminary estimate of $\vect{u}$, which is then used to construct a more precise weighting matrix $\mat{W} = \left( \mat{B}\mat{Q}\mat{B}^{T} \right)^{-1}$. Finally, solving the estimation problem again with this updated matrix yields a refined and more accurate solution.
	
	\section{Cramér-Rao Lower Bound}
	In estimation theory, the Cramér-Rao Lower Bound (CRLB) provides a theoretical limit on the variance achievable by any unbiased estimator. Under the assumption of Gaussian-distributed noise, this lower bound for the source position, $\widetilde{\vect{u}}^{o} = [\varphi^{o},\theta^{o},g^{o}]^{T}$ in MPR is derived in \cite{kay1993} as
	\begin{equation}
	\text{CRLB}\left( \widetilde{\vect{u}}^{o} \right) = \left( \frac{\partial{\vect{\psi}^{o}}^{T}}{\partial\widetilde{\vect{u}}^{o}}\mat{Q}^{-1}\frac{\partial\vect{\psi}^{o}}{\partial{\widetilde{\vect{u}}^{oT}}} \right)^{-1}
	\end{equation}
	To simplify the analysis, let us define the vector $\vect{\omega}^{o} = [{{\vect{\rho}}^{o}}^{T},g^{o}]^{T}$ as an intermediate variable to facilitate the computation
	\begin{equation}
	\text{CRLB}\left( \widetilde{\vect{u}}^{o} \right) = \mat{D} \text{CRLB}\left( \vect{\omega}^{o} \right) \mat{D}^{T}
	\end{equation}
	where $\mat{D}$ is the derivation of $\widetilde{\vect{u}}^{o}$ with respect to $\vect{\omega}^{o}$ and is given by
	\begin{equation}
	\mat{D} = \begin{bmatrix}
	\frac{-\sin(\varphi^{o})}{\cos(\theta^{o})} & \frac{\cos(\varphi^{o})}{\cos(\theta^{o})} & 0 & 0 \\
	{- \cos}(\varphi^{o})\sin(\theta^{o}) & {- \sin}(\varphi^{o})\sin(\theta^{o}) & \cos(\theta^{o}) & 0 \\
	0 & 0 & 0 & 1
	\end{bmatrix}
	\end{equation}
	The CRLB for $\vect{\omega}^{o}$ can be written as
	\begin{equation}
	\text{CRLB}\left( \vect{\omega}^{o} \right) = \left( \frac{\partial{\vect{\psi}^{o}}^{T}}{\partial\vect{\omega}^{o}}\mat{Q}^{-1}\frac{\partial\vect{\psi}^{o}}{\partial{\vect{\omega}^{o}}^{T}} \right)^{-1}
	\end{equation}
	The partial derivation of $\vect{\psi}^{o}$ with respect to $\vect{\omega}^{o}$ is
	\begin{equation}
	\frac{\partial\vect{\psi}^{o}}{\partial{\vect{\omega}^{o}}^{T}} = \left[ \frac{\partial{\vect{\psi}^{o}}^{T}}{\partial{{\vect{\rho}}}^{o}},\frac{\partial{\vect{\psi}^{o}}^{T}}{\partial g} \right]^{T}
	\end{equation}
	where
	\begin{subequations}
	\begin{align}
	&\left[ \frac{\partial\vect{\psi}^{o}}{\partial{{{\vect{\rho}}}^{o}}^{T}} \right]_{i,:} = - \frac{\left\| {\vect{\rho}}^{o} - \vect{s}_{i}^{o}g^{o} \right\|^{2} {\vect{a}_{i}^{o}}^{T} + g^{o}({\vect{a}_{i}^{o}}^{T}\left( {\vect{\rho}}^{o} - \vect{s}_{i}^{o}g^{o} \right)){\vect{s}_{i}^{o}}^{T}}{\left\| {\vect{\rho}}^{o} - \vect{s}_{i}^{o}g^{o} \right\|^{3}\left| \sin\left( \psi_{i}^{o} \right) \right|} \\
	&\left[ \frac{\partial\vect{\psi}^{o}}{\partial g} \right]_i =
	\frac{1}{\left\| {\vect{\rho}}^{o} - \vect{s}_{i}^{o}g^{o} \right\|^{3}\left| \sin\left( \psi_{i}^{o} \right) \right|}({\vect{a}_{i}^{o}}^{T}\vect{s}_{i}^{o}\left\| {\vect{\rho}}^{o} - \vect{s}_{i}^{o}g^{o} \right\|^{2} + \nonumber\\
	& ({\vect{a}_{i}^{o}}^{T}\left( {\vect{\rho}}^{o} - \vect{s}_{i}^{o}g^{o} \right))(g^o\|{\vect{s}_{i}^{o}}\|^2-{\vect{s}_{i}^{oT}}\boldsymbol{\rho}^o)) , i=1,2,\ldots,N\nonumber
	\end{align}
		\end{subequations}
	
	\section{Simulation Results}
	This section of the paper presents a performance evaluation of the proposed estimator via numerical simulations. For comparison purpose, we have included the 2SDP \cite{alamdari2022}, ReSDP \cite{sun2022}, CWLS-C \cite{li2025} and the MLE. The MLE is implemented using the Gauss-Newton algorithm with the true values for azimuth, elevation and the inverse-range of the source as the initial guess. The CVX toolbox \cite{grant2008} is employed to solve the SDP in \eqref{eq:Problem4}, \cite{alamdari2022} and \cite{sun2022}.
	
	The simulations were carried out in a 3-D space employing a network of 12 sensors. In both scenarios, 10 random geometries were generated, with 1000 measurements being produced for each geometry to compute the average MSE. In each network configuration, sensors are placed randomly in a $500\times500\times500$ cube centered at $[0,0,0]^{T}$ and their azimuth and elevation angle are randomly chosen using uniform distribution $\mathcal{U}(-\pi,\pi)$ and $\mathcal{U}\left( \frac{-\pi}{2},\frac{\pi}{2} \right)$, respectively. The 1-D angle measurements are generated according to \eqref{eq:meas} and the measurement error follows zero mean Gaussian model with diagonal covariance matrix $\mat{Q}$ where the diagonal elements are randomly generated and the averaged measurement noise power satisfies $\frac{\text{tr}(\mat{Q})}{M} = \sigma_{m}^{2}$. The MSE is used to evaluate the performance of the propose method by $\text{MSE}(\varphi,\theta) = \frac{1}{NL}\sum_{j = 1}^{N}{\sum_{l = 1}^{L}\left[ \left( \varphi_{j}^{o} - \varphi_{jl} \right)^{2} + \left( \theta_{j}^{o} - \theta_{jl} \right)^{2} \right]}$ and $\text{MSE}(g) = \frac{1}{NL}\sum_{j = 1}^{N}{\sum_{l = 1}^{L}\left( g_{j}^{o} - g_{jl} \right)^{2}}$ where $\varphi_{jl}$, $\theta_{jl}$ and $g_{jl}$ are the estimates of $\varphi_{j}^{o}$, $\theta_{j}^{o}$ and $g_{j}^{o}$ at ensemble run $l$ in the sensor geometry $j$, respectively.
	
	\subsection{Scenario 1: Near-Field Localization}
	This scenario evaluates the near-field point positioning performance of the proposed method under increasing measurement noise power. The source range is fixed at 1000 m, with its azimuth and elevation angles randomly generated for each trial using a uniform distribution. The MSE performance for the angle and inverse-range estimates is depicted in Figs. 1 and 2, respectively. The proposed method attains the CRLB when the noise power does not exceed 0.00316 rad². For angle estimation, its performance is comparable to that of the 2SDP and ReSDP methods. However, for inverse-range estimation, the proposed method demonstrates an accuracy improvement of approximately 6 dB over these two methods when the noise power exceeds 0.00316 rad². In contrast, the CWLS-C method begins to exhibit a significant deviation from the CRLB when the noise power surpasses 0.000316 rad².
	
	\begin{figure}[htbp]
		\centering
		\includegraphics[width=0.45\textwidth]{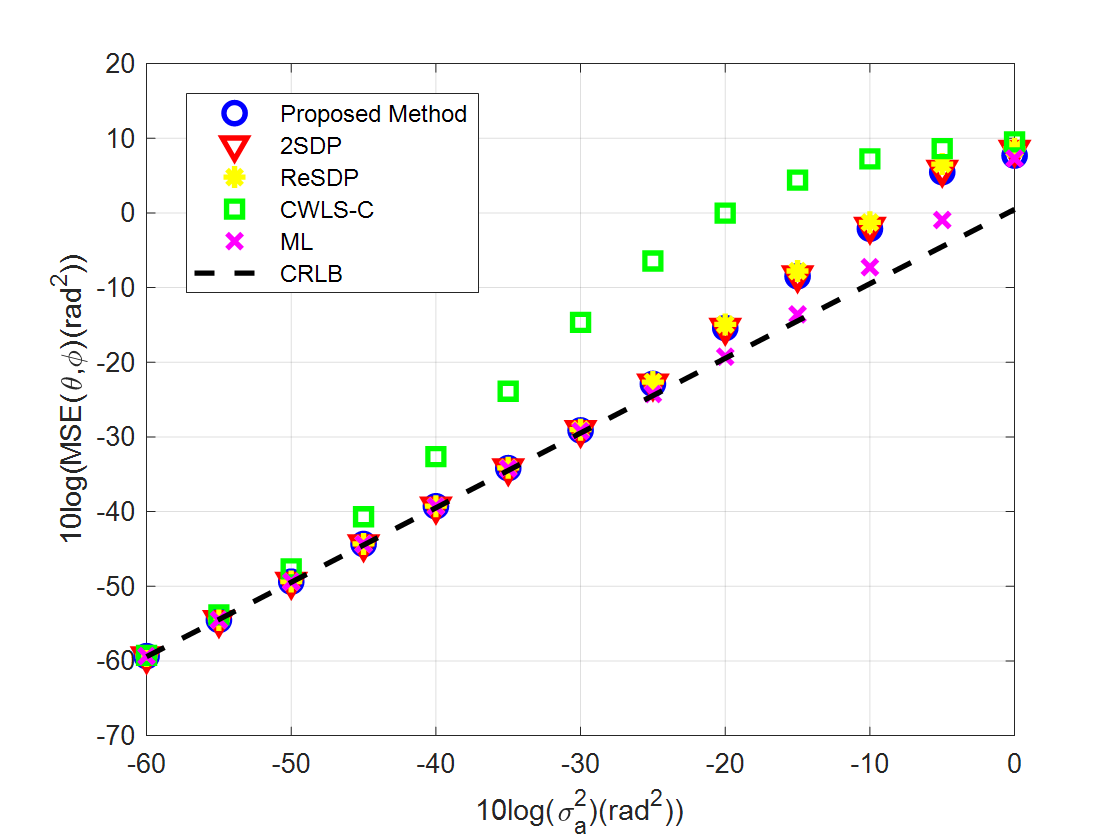}
		\caption{MSE performance of the angle estimation as a function of measurement noise variance.}
		\label{fig:angle_near}
	\end{figure}
	
	\begin{figure}[htbp]
		\centering
		\includegraphics[width=0.45\textwidth]{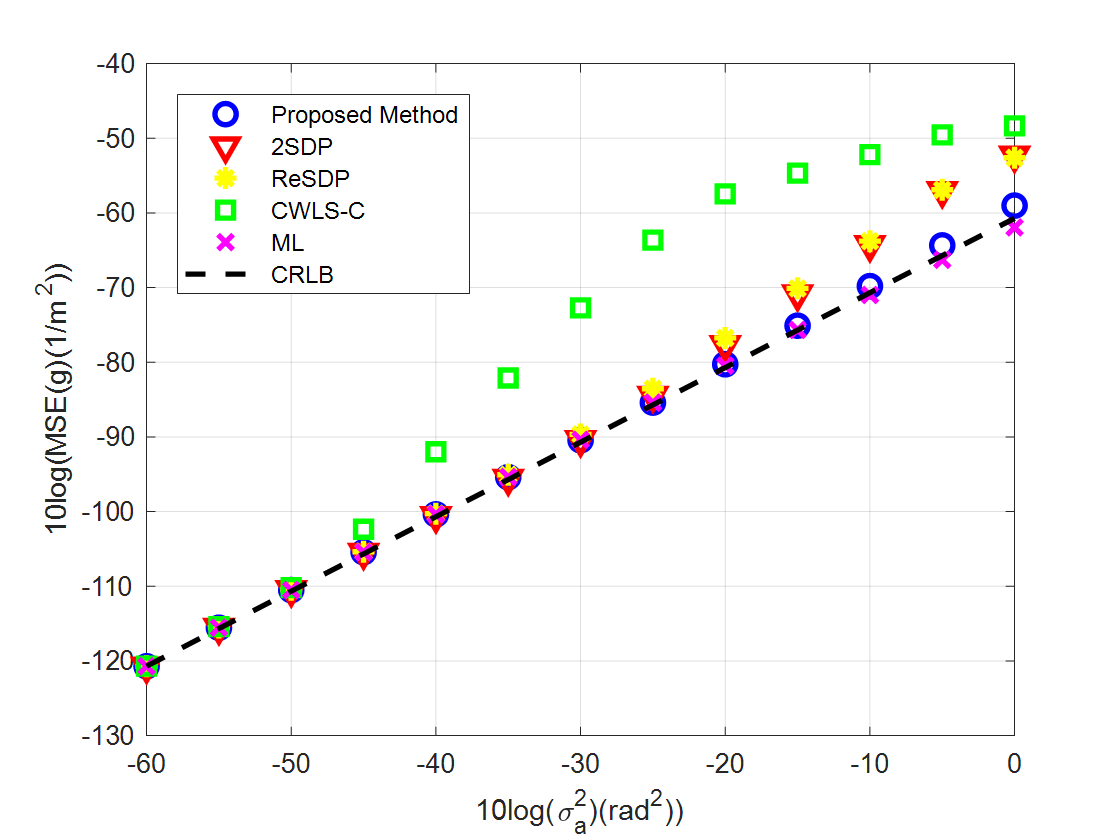}
		\caption{MSE performance of the inverse-range estimation as a function of measurement noise variance.}
		\label{fig:inverse_range_near}
	\end{figure}
	
	\subsection{Scenario 2: Far-Field Localization}
	The core objective of this paper was to design an estimator that maintains optimal performance regardless of the source range. As shown in Figs. 3-4, the proposed method's superiority becomes evident with increasing source range. While Cartesian SDP estimators 2SDP and ReSDP begin to deviate from the CRLB beyond 5 km and 10 km, respectively, the CWLS-C solution shows significant deviation even at moderate ranges. This confirms the proposed estimator's robustness and superior far-field performance.
	
	\begin{figure}[htbp]
		\centering
		\includegraphics[width=0.45\textwidth]{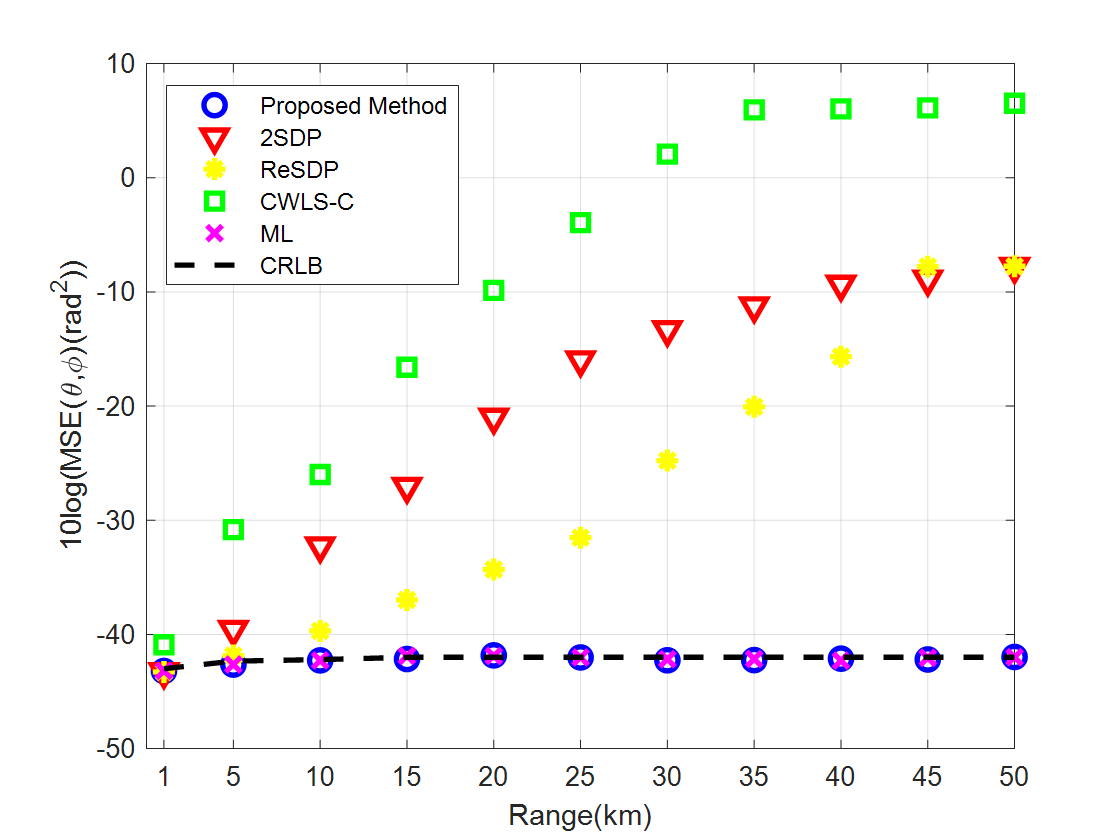}
		\caption{MSE performance of the angle estimation as a function of source range.}
		\label{fig:angle_far}
	\end{figure}
	
	\begin{figure}[htbp]
		\centering
		\includegraphics[width=0.45\textwidth]{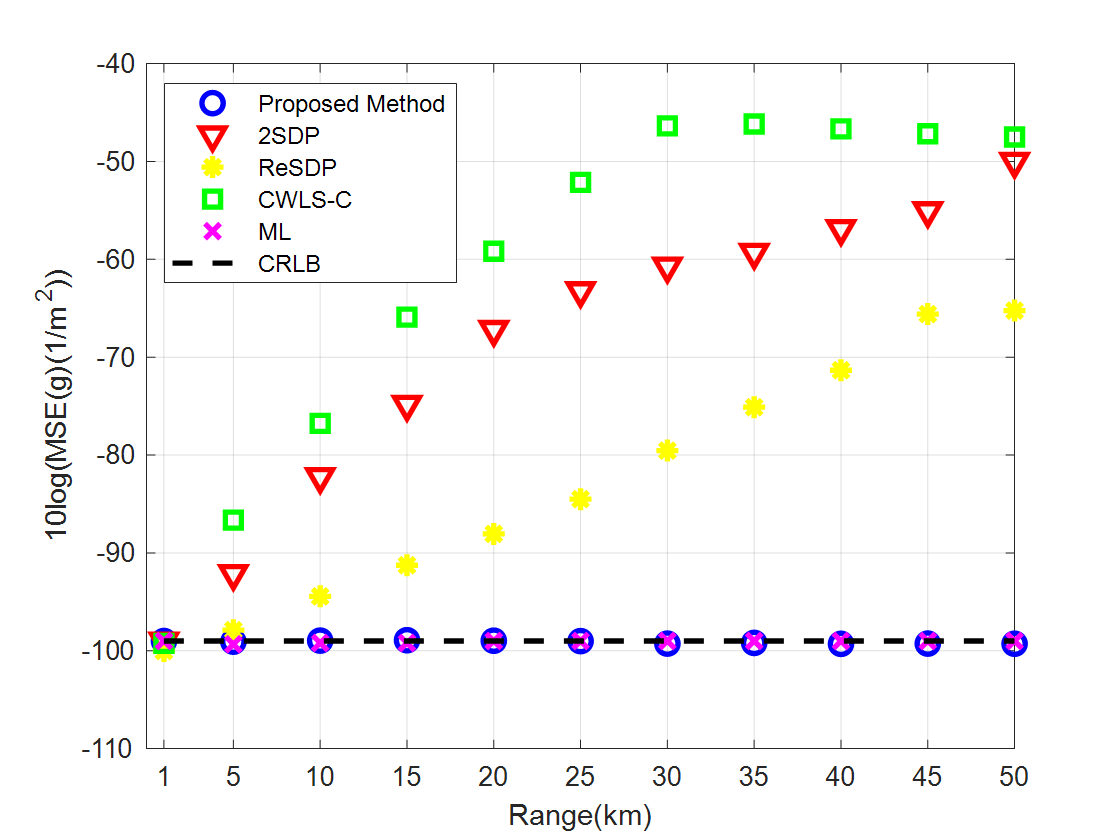}
		\caption{MSE performance of the inverse-range estimation as a function of source range.}
		\label{fig:inverse_range_far}
	\end{figure}
	
	\section{Conclusion}
	In this paper, we addressed the critical challenge of far-field performance degradation in 3-D source localization using networks of one-dimensional arrays. By reformulating the problem within the MPR coordinate system and developing a novel constrained SDP estimator, we have established a unified framework that remains robust irrespective of the source range. The proposed method effectively overcomes the limitations of existing Cartesian-based estimators, which exhibit significant performance deterioration as the source distance increases. Simulation results demonstrate that our estimator attains the CRLB for both angle and inverse-range parameters in near-field scenarios, while maintaining superior accuracy in the far-field, thereby fulfilling the core objective of reliable and unified localization.
	
\bibliographystyle{IEEEtran}
\balance
\bibliography{refs}
	
\end{document}